\documentclass{elsart}

 \usepackage{graphicx}
\usepackage{amssymb}

\begin{document}

\begin{frontmatter}

% Title, authors and addresses

% use the thanksref command within \title, \author or \address for footnotes;
% use the corauthref command within \author for corresponding author footnotes;
% use the ead command for the email address,
% and the form \ead[url] for the home page:
% \title{Title\thanksref{label1}}
% \thanks[label1]{}
% \author{Name\corauthref{cor1}\thanksref{label2}}
% \ead{email address}
% \ead[url]{home page}
% \thanks[label2]{}
% \corauth[cor1]{}
% \address{Address\thanksref{label3}}
% \thanks[label3]{}

%\title{Spin-powered very high energy emission from the BL Lac \\ 1ES 0806+524}
\title{Magnetocentrifugal mechanism of pair creation in AGN}

% use optional labels to link authors explicitly to addresses:
% \author[label1,label2]{}
% \address[label1]{}
% \address[label2]{}

\author{Zaza N. Osmanov} 

\address{School of Physics, Free University of Tbilisi, 0183, Tbilisi, Georgia}

\address{E. Kharadze Georgian National Astrophysical Observatory, Abastumani 0301, Georgia}

\author{Gianluigi Bodo \& Paola Rossi} 

\address{INAF/Osservatorio Astrofisico di Torino, Strada Osservatorio 20, 10025 Pino Torinese, Italy}

\begin{abstract}
In the manuscript, we study the efficiency of pair creation by means of the centrifugal mechanism. The strong magnetic field and the effects of rotation, which always take place in Kerr-type black holes, guarantee the frozen-in condition, leading to the generation of an exponentially amplifying electrostatic field. This field, when reaching the Schwinger threshold, leads to efficient pair production. The process has been studied for a wide range of AGN luminosities and black hole masses, and it was found that the mechanism is very efficient, indicating that for AGNs where centrifugal effects are significant, the annihilation lines in the MeV range will be very strong.
\end{abstract}


\end{frontmatter}

% main text

\section{Introduction}

In the literature, the population of electron-positron pairs in AGN magnetospheres has been studied from different perspectives. In the framework of Penrose pair production, the MeV photons originating in the inner accretion disk and entering the ergosphere may increase their energy via the blueshift effect. In due course, the energy will reach the GeV threshold, which is enough for pair creation after scattering off the protons \cite{penrose}. Another popular scenario is the so-called $\gamma \gamma$ process, when high-energy photons scatter off relatively soft photons, always present in the accretion disks, and produce electron-positron pairs \cite{LZ}. As it has been found, these are not the only mechanisms providing the population of $e^+e^-$ pairs. The present paper is dedicated to the study of the new mechanism of pair production in the AGN magnetospheres.

In a recent paper \cite{paircr}, a new mechanism of pair creation in the magnetospheres of pulsars has been presented. In particular, it has been shown that since the magnetospheres of pulsars are characterized by rotation, magnetocentrifugal effects might lead to the generation of Langmuir waves. On the other hand, the centrifugal force harmonically depends on time \cite{gedank}, leading to the parametric instability of the process and thus to the exponential growth of the electrostatic field. By means of this growth, under certain conditions, the electrostatic field will approach the Schwinger threshold, $E_S = \pi m^2c^3/(e\hbar)\simeq 1.4\times 10^{14}$ statvolt cm$^{-1}$ \cite{heisen,sauter,schwinger}, when pair creation might start. Here $m$ and $e$ are the electron's mass and charge respectively, $c$ is the speed of light; and $\hbar$ denotes the Planck constant. In particular, quantum electrodynamics considers the vacuum as a complex system composed of virtual particles and antiparticles continuously creating and annihilating. The strong electric field ($E$) on the other hand, if its work done on the Compton wave-length, $\lambda_{_C}$, is of the order of the necessary energy of materialised pairs, $eE\lambda_{_C}\simeq 2mc^2$, will lead to extremely efficient pair creation. 

\begin{figure}
  \resizebox{\hsize}{!}{\includegraphics[angle=0]{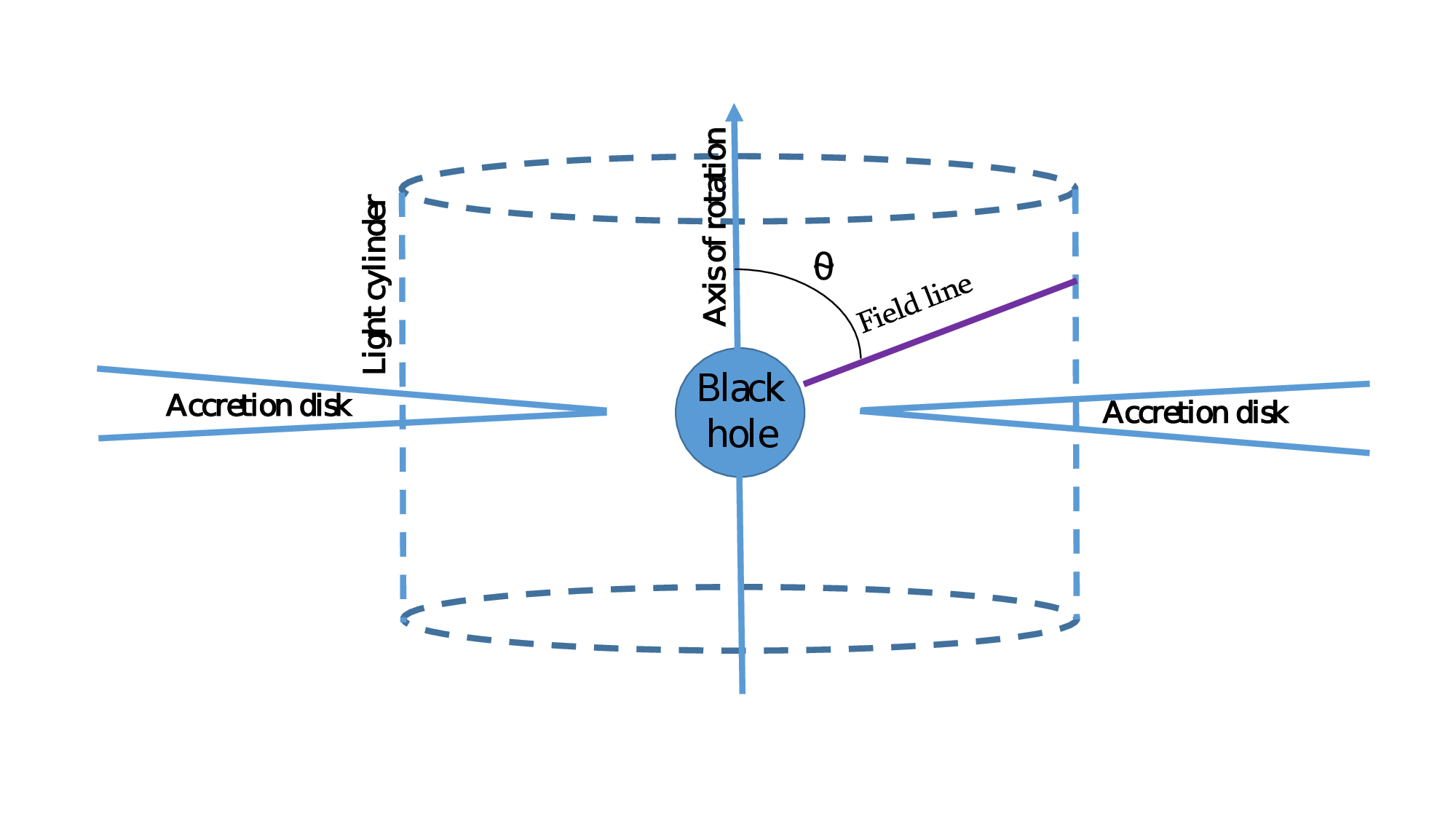}}
  \caption{Sketch of the model, with the centrifugally accelerated co-rotating particles in the nearby zone of the LC area of the Kerr-type black hole with an accretion disk.}\label{fig1}
\end{figure}

It is generally accepted that plasma particles in a nearby region of AGN are embedded in the magnetic field strong enough to provide the frozen in-condition \cite{BZ}, which when combined with effects of rotation (always present in Kerr-type black holes \cite{carroll,shapiro}) leads to the relativistic magnetocentrifugal effects close to the light cylinder zone \cite{gold} - a hypothetical area where the linear velocity of rotation exactly equals the speed of light (See the sketch in Fig. 1). As it has been shown, the magnetocentrifugal acceleration might play an important role in particle energization in AGN \cite{bodo,blazar}, when particles might reach energies of the order of $10$TeV. This very effect will inevitably induce Langmuir waves in the magnetospheres of back holes as well.

A series of papers \cite{zev,pev,jet} has been dedicated to studying this particular problem in AGN (also considering particle acceleration) and it was found that the magnetocentrifugal generation of Langmuir waves was so efficient that the energy pumped by the electrostatic modes from rotation was enormous. A similar study for pulsars also showed an extremely efficient character of the excitation of electrostatic modes \cite{srep1,srep2}

In the framework of these works, an approximate expression of GJ particle number density has been used \cite{GJ} without taking general relativistic effects into account. This particular problem has been studied in \cite{rieger} where the authors have examined Kerr-type black holes and derived a general-relativistic expression of GJ density.

In \cite{LangAGN} the excitation of magnetocentrifugally driven Langmuir waves has been explored by taking the general expression of GJ density into account. The increment of the process has been studied for a wide range of physical parameters, including the particle Lorentz factors, AGN luminosity, and the mass of the central object. It was found that the time scale of the process is small compared to the kinematic time scale of rotation, indicating high efficiency of the exponential amplification of the electric field.

The exponentially increasingg electric field will reach the Schwinger threshold, and pair production will start. In the present manuscript, we examine the new mechanism of the magnetocentrifugally driven pair creation in the AGN magnetospheres and explore the process versus important physical parameters.

The paper is organized as follows: in Sec.2, the general framework of the approach will be outlined; in Sec. 3, we will apply the model to AGN, obtaining results; and in Sec. 4, we summarize them.

%%%%%%%%%%%%%%%%%%%%%%%%%%%%%%%%%%%%%%%%%%
\section{The Framework}
In this section, we briefly outline the theoretical model of centrifugally excited Langmuir waves and the corresponding pair cascading that takes place when the electric field approaches the Schwinger threshold. In the framework of the paper, we assume that the magnetic fields are almost straight because, as the study shows \cite{twist}, they change their rectilinear configuration on the LC surface, and the particles most of their evolution follows the unperturbed field lines.

The generation of electrostatic waves is fully governed by the set of following equations (rewritten in Fourier space) \cite{LangAGN}
 \begin{equation}
\label{eul} \frac{\partial p_{_{\beta}}}{\partial
t}+ik\upsilon_{_{\beta0}}p_{_{\beta}}=
\upsilon_{_{\beta0}}\Omega_0^2r_{_{\beta}}p_{_{\beta}}+\frac{e_{_{\beta}}}{m_{_{\beta}}}E,
\end{equation}
\begin{equation}
\label{cont} \frac{\partial n_{_{\beta}}}{\partial
t}+ik\upsilon_{_{\beta0}}n_{_{\beta}}, +
ikn_{_{\beta0}}\upsilon_{_{\beta}}=0
\end{equation}
\begin{equation}
\label{pois} ikE=4\pi\sum_{_{\beta}}n_{_{\beta0}}e_{_{\beta}},
\end{equation}
where Eq. (\ref{eul}) represents the Euler equation, Eq. (\ref{cont}) is the continuity equation and Eq. (\ref{pois}) is the Poisson equation. We used the following notations: $p_{_{\beta}}$ denotes the dimensionless first order momentum,  ${\beta}$ is an index of species (protons or electrons), $k$ represents the wave number, $\upsilon_{_{\beta0}}(t) \approx
c\cos\left(\Omega_0 t + \phi_{_{\beta}}\right)$ is the unperturbed
velocity, $\Omega_0$ denotes the angular momentum of rotation, $\phi_{_{\beta}}$ represents a phase $r_{_{\beta}}(t) \approx
\frac{c}{\Omega_0}\sin\left(\Omega_0 t + \phi_{_{\beta}}\right)$ is the
radial coordinate of a corresponding species, $e_{_{\beta}}$ denotes
charge and $n_{_{\beta}}$ and $n_{_{\beta0}}$ are respectively the first and the zeroth order Fourier terms of the number density. 

Following the method originally developed for AGN in \cite{jet} (see also the detailed study in \cite{LangAGN}) one obtains the dispersion relation of the process
\begin{equation}
\label{disp} \omega^2 -\omega_e^2 - \omega_p^2  J_0^2(b)= \omega_p^2
\sum_{\mu} J_{\mu}^{2}(b) \frac{\omega^2}{(\omega-\mu\Omega_0)^2},
\end{equation}
leading to the growth rate of the instability
\begin{equation}
 \label{grow1}
 \Gamma= \frac{\sqrt3}{2}\left (\frac{\omega_e {\omega_p}^2}{2}\right)^{\frac{1}{3}}
 {J_{\mu}(b)}^{\frac{2}{3}},
\end{equation}
where $\omega$ denotes the frequency of the electrostatic waves, $\omega_{e,p}\equiv\sqrt{4\pi e^2n_{e,p}/m_{e,p}\gamma_{e,p}^3}$ represents the plasma frequency of a corresponding specie (electrons and protons), $n_{e,p}$, $m_{e,p}$ and $\gamma_{e,p}$ are respectively the density, mass, and the relativistic factor of the mentioned components, $J_{\nu}(x)$ is the Bessel function of the first kind, $b=2ck\sin\phi/\Omega_0$ and $\mu = \omega_e/\Omega_0$.

The evolution of the electric field then writes as
\begin{equation}
\label{E} 
E = E_0 e^{\Gamma t},
\end{equation}
where for the initial value of the electric field, one can use the Gauss's law
\begin{equation}
\label{E0} 
E_0\simeq 4\pi n\Delta r,
\end{equation}
where $n$ is the number density of particles and $\Delta r = \frac{\gamma}{d\gamma/dr}$ is a spatial scale where the centrifugal effects are supposed to be most important. By taking an expression of the Lorentz factor of centrifugally accelerated particles into account \cite{rieg}
\begin{equation}
\label{gamma} 
\gamma = \frac{\gamma_0}{1-r^2/R_{lc}^2},
\end{equation}
one can estimate the scale of the shell $\Delta r\simeq \gamma_0R_{lc}/(2\gamma)$ \cite{LangAGN}. Here $\gamma_0$ is the initial relativistic factor and $R_{lc} = c/\Omega_0$ represents the LC radius.

If the particle distribution is detrmined by rotation, then the particle density should equal the GJ density, which in the general-relativistic scenario is given by \cite{rieger}
\begin{equation}
\label{GJ} 
n_{_{GJ}}\simeq \frac{\left(\Omega-\Omega^F\right)B_Hr_H^2\cos\theta}{\pi ce\alpha_l\rho^2},
\end{equation}
where $\Omega = 2c\alpha_s r_gr/\Sigma^2$ is the angular velocity with respect to the absolute space, $\alpha_s = a r_g$, $r_g = GM/c^2$, $M$ represents the black hole mass, $\Omega^F = c\alpha_s/4r_gr_H$ denotes the angular velocity by which the magnetic field lines co-rotate, $r_H = r_g+\sqrt{r_g^2-\alpha_s^2}$ is the event horizon radius, $\alpha_l = \rho\sqrt{\Delta}/\Sigma$, $\rho^2 = r^2+\alpha_s^2\cos^2\theta$, $\Delta = r^2-2rr_g+\alpha_s^2$ and $\Sigma^2 = \left(r^2+\alpha_s^2\right)^2-\alpha_s^2\Delta\sin^2\theta$ and $\theta$ is the angle relative to the axis of rotation.

By means of the instability of the centrifugally induced Langmuir waves, the electric field will reach the Schwinger threshold. For high values of the electric field, the pair creation rate per unit of volume is given by \cite{schwinger,MP}
\begin{equation}
 \label{rate}
 R\equiv\frac{dN}{dtdV} = \frac{e^2E^2}{4\pi^3c\hbar^2}\sum_{k}\frac{1}{k^2}exp\left({-\frac{\pi m^2c^3}{e\hbar E}k}\right).
\end{equation}
This expression is valid for constant electric fields. One can straightforwardly check that the Langmuir frequency is of the order of $10^{4-5}$ Hz. On the other hand, the characteristic frequency of par creation $\nu\simeq 2mc^2/h\sim 10^{20}$ Hz exceeds by many order of magnitude the plasma frequency, indicating that the aforementioned expression realistically describes the pair cascading process. As it is clear from Eq. (\ref{rate}) the pair creation will occur not only for the Schwinger threshold, when the process becomes extremely efficient, but also for values less than $E_S$. On the other hand, if the difference between $E_S$ and $E$ becomes large, the process will be exponentially suppressed.

%%%%%%%%%%%%%%%%%%%%%%%%%%%%%%%%%%%%%%%%%%
\section{Discussion and Results}

The Kerr-type black holes are rotating with the angular velocity \cite{shapiro}
\begin{equation}
\label{rotat} \Omega\approx\frac{a c^3}{2GM\left(1+\sqrt{1-a^2}\right)}\approx
2.5\times 10^{-2}\frac{a}{M_8\left(1+\sqrt{1-a^2}\right)}rad/s^2,
\end{equation}
where $M_8 = M/(10^8M_{\odot})$ is a dimensionless mass parameter of the black hole and $M_{\odot}\simeq 2\times 10^{33}$ g is the solar mass. Rotation combined with the magnetic field will lead to the magnetocentrifugal process of acceleration.

In the framework of the paper, we assume the equipartition approximation when the magnetic field energy density is of the order of the AGN emission energy density and when for the magnetic induction one obtains \begin{equation}
 \label{B}
B\simeq \frac{1}{r}\sqrt{\frac{2L}{c}}\simeq 2.8\times 10^2\times \frac{R_H}{r}\times L_{42}^{1/2},
 \end{equation}
where $L$ is the luminosity of AGN and we use the dimensionless luminosity $L_{42}=L/(10^{42} erg\; s^{-1})$. In such a strong magnetic field, electrons have a gyro-radius of the order of $R_{gyro} \simeq\gamma mc/(eB)\simeq 10^{-5}\times\gamma/10^4$ cm, which is by many orders of magnitude less than the spacial scale factor of the process - LC radius, therefore, the plasma particles are in frozen-in condition and particles centrifugally accelerate. But the acceleration process might be limited by several factors.
 
Moving in the strong magnetic field, the particles will experience very efficient synchrotron losses with the power $P_{s}\simeq 2e^4B^2\gamma^2/(3m_pc^3)$. Then, for the corresponding time scale of energy losses, one obtains $\gamma m_pc^2/P_s$ which for the same Lorentz factor, $10^4$, is much less than the rotation period of the black hole's nearby zone, $2\pi/\Omega$. As a result, the particles, soon after they start accelerating, lose their perpendicular momentum, transit to the ground Landau state, and continue sliding along the field lines. Therefore, the synchrotron mechanism does not impose any constraints on particle energies. A similar scenario takes place for protons as well.

Another constraint has been introduced in \cite{rieg} for the field lines co-rotating in the equatorial place and developed in \cite{bodo} for the inclined ones: the particles experience an effective reaction force from the field lines. On the other hand, the same charged particles experience the magnetic Lorentz force. Initially the particles follow the field lines, but in due course of time the reaction force will exceed the Lorentz one (on the LC area) violating the frozen-in condition and for the maximum Lorentz factor one obtains 
 
\begin{equation}
\label{BBW1} 
\gamma_{max}^{BBW}\simeq A_1+\left[A_2+\left(A_2^2-A_1^6\right)^{1/2}\right]^{1/3}+\left[A_2-\left(A_2^2-A_1^6\right)^{1/2}\right]^{1/3},
\end{equation}
with
\begin{equation}
\label{A1} A_1 = -\frac{\gamma_0ctg^2\theta}{12},
\end{equation}
\begin{equation}
\label{A2} A_2 = \frac{\gamma_0e^2B_{lc}^2}{4m^2c^2\Omega^2}+A_1^3,
\end{equation}
where $B_{lc}$ is the magnetic field on the LC length-scales.

The particles moving in a photon sea will lead to another process: the inverse Compton scattering, when the accelerated particles encounter soft photons. Saturation occurs when the energy gain and the cooling process balance each other. When this happens, the maximum relativistic factor achievable by electrons writes as \cite{bodo}
\begin{equation}
\label{IC} 
\gamma_{max}^{IC}\simeq\left(\frac{8\pi m_ec^4}{\gamma_0\sigma_TL\Omega}\right)^2,
\end{equation}
where $\sigma_T$ denotes the Thomson cross section. For protons, the same mechanism is strongly suppressed \cite{ahar}, therefore, in the aforementioned expression, we used the electron's mass.

When charged particles move on curved trajectories, they emit curvature radiation, which, when balanced with the energy gain due to the centrifugal acceleration, achieves the maximum Lorentz factor \cite{review}
\begin{equation}
\label{gcur} 
\gamma_{max}^c\simeq\frac{1}{\gamma_0^{1/5}}\times\left(\frac{3\pi mc^3\sin\alpha}{2\pi  e^2\Omega}\right)^{2/5}\times\left(\frac{R_c}{R_{lc}}\right)^{4/5},
\end{equation} 
where $R_c$ is the curvature radius of the trajectory, and we assume that the trajectory close to the LC is almost circle \cite{gold} and therefore, $R_c\simeq R_{lc}$.

It is clear that the mechanism that provides the minimum value of the relativistic factor is a leading process in limiting the maximum achievable energies. One can straightforwardly check that if $\theta\simeq\pi/2$ for electrons, this is the IC scattering with  
\begin{equation}
\label{ge} 
\gamma_{e,max}\simeq 1.2\times 10^4\times\left(\frac{10}{\gamma_0}\times \frac{M_8}{L_{42}}\right)^{2}
\end{equation} 
and for protons - the BBW process
\begin{equation}
\label{gp} 
\gamma_{p,max}\simeq 3.2\times 10^6\times M_8^{2/3}\times\left(L_{42}\times\frac{\gamma_0}{10}\right)^{1/3}.
\end{equation} 

\begin{figure}
  \resizebox{\hsize}{!}{\includegraphics[angle=0]{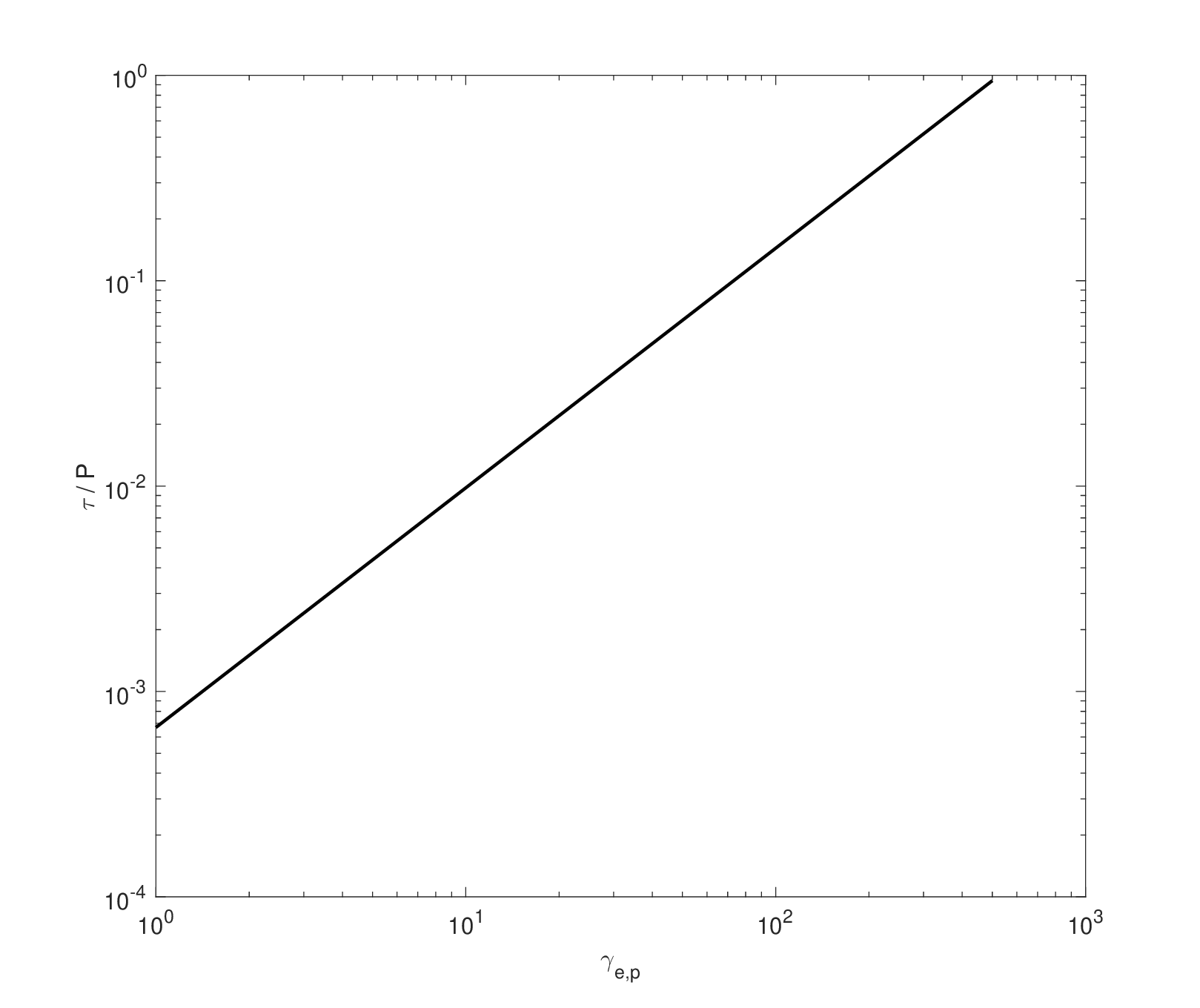}}
  \caption{Behaviour of the dimensionless time-scale versus $\gamma_{e,p}$. The set of parameters is: $M_8 = 1$, $L_{42}=1$, $\theta\simeq 89^0$, $r = R_{lc}/\sin\theta$, $\gamma_0 = 1$.}\label{fig2}
\end{figure}

\begin{figure}
  \resizebox{\hsize}{!}{\includegraphics[angle=0]{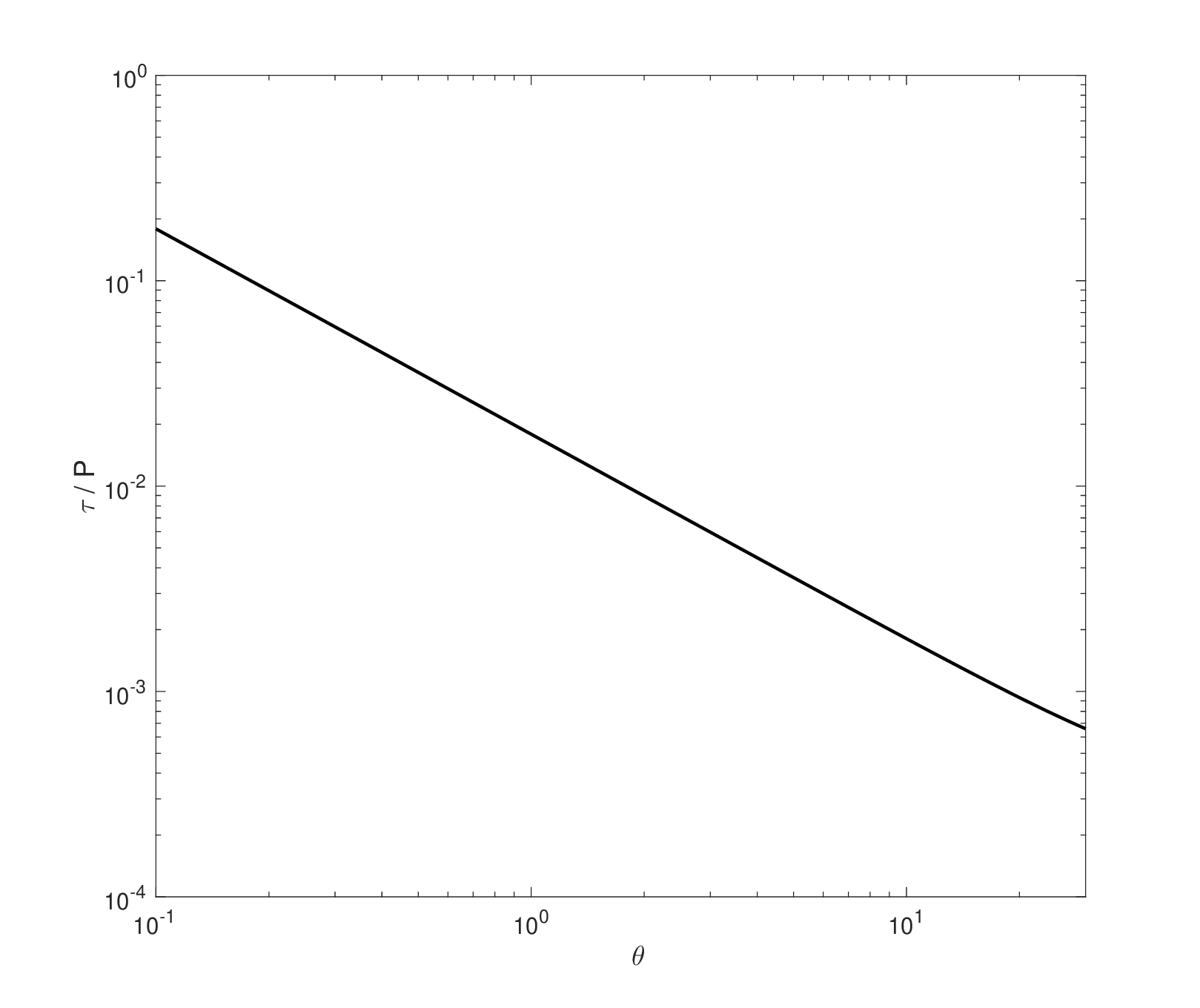}}
  \caption{Behaviour of the dimensionless time-scale versus the inclination angle of the field line relative to the rotation axis. The set of parameters is: $M_8 = 1$, $L_{42}=1$, $r = R_{lc}/\sin\theta$, $\gamma_0 = 1$, $\gamma_{e,p} = 2$.}\label{fig3}
\end{figure}

In Fig. 2 we plot the instability time scale (normalised by the rotation period of the black hole, $P = 2\pi/\Omega$) versus the Lorentz factors, which are supposed to be equal. The set of parameters is: $M_8 = 1$, $L_{42}=1$, $\theta\simeq 89^0$, $r = R_{lc}/\sin\theta$, $\gamma_0 = 1$. As it is evident from the figure, for almost the whole range of the considered Lorentz factors, the instability timescale is small compared to the period of rotation, indicating high efficiency of the process. The higher efficiency comes for the smallest Lorrentz factors, but on the other hand, the instability strongly depends on the relativistic character of the process: the radial velocity behaves as $\upsilon\simeq c\cos\left(\Omega_0 t+\phi\right)$ \cite{gedank}, therefore, one could assume $\gamma_{e,p}\gtrsim 2$.

Due to the instability, the electric field exponentially increases, and by approaching the Schwinger threshold, pair creation will initiate (see Eq. (\ref{rate})). It is clear from Eq. (\ref{rate}), when it is started, the rate becomes very high. On the other hand, by creating the pairs, the plasma energy density will be significantly increased, which will have a feedback effect on the process itself. In particular, the energy balance leads to a condition when the pair plasma power density gain becomes of the order of the electric power density\begin{equation}
\label{term1} 
2m_ec^2R(t)\simeq\frac{d}{dt}\left(\frac{E^2(t)}{8\pi}\right).
\end{equation} 
This is an algebraic equation for solving a time-scale $t_0$, when the condition is satisfied. By combining Eqs.(\ref{E},\ref{rate}), for $\gamma_{e,p} = 10$ the aforementioned expression leads to $E\simeq E_0e^{\Gamma t_0}\simeq 3\times 10^{12}$ statvolt cm$^{-1}$. With such a high value, the pair production rate becomes quite high. From Eq. (\ref{rate}) one can show that $R\simeq 2\times 10^{28}$ cm$^{-3}$ s$^{-1}$. 

The corresponding time-scale when the pair production rate reached this value equals $t_0\simeq10^4$ sec, which for the number density gives $n_{pair}\simeq Rt_0\simeq 10^{32}$ cm$^{-3}$. This in turn leads to an annihilation time-scale of the order of $\tau_{ann}\simeq 1/(\sigma n_{pair}c)\simeq 10^{-19}$ sec ($\sigma\simeq 10^{-24}$ cm$^{-3}$ is the Thomson cross section). Therefore, in limiting the pair production rate, the annihilation process has to be taken into account.

In particular, for non-relativistic temperatures, it has been found that the annihilation rate is given by \cite{ann} 
\begin{equation}
\label{ann} 
\Lambda\simeq 2\pi c r_e^2 n_-n_+,
\end{equation} 
where $n_-$ and $n_+$ are the number densities of electrons and positrons, respectively, and $r_e$ is the electron's classical radius. By taking the natural relation $n_-=n_+$ into account, the balance between the production and the annihilation processes
\begin{equation}
\label{bal} 
R(\tau)\simeq\Lambda\simeq 2\pi c r_e^2 \left(\int_0^{\tau}R(t)dt\right)^2,
\end{equation} 
which, after taking the derivative (by $\tau$) of Eq. (\ref{bal}) and neglecting the term $e^{\Gamma\tau}$ compared to $e^{2\Gamma\tau}$ in the left hand side of equation, straightforwardly leads to an estimate of densities of electrons and positrons $\int_0^{\tau}R(t)dt$ 
\begin{equation}
\label{dens} 
n_{\pm}\simeq\frac{\Gamma}{2 \pi c r_e^2}\simeq 1.3\times 10^{11} cm^{-3},
\end{equation} 
which for the pair creation rate gives the value
\begin{equation}
\label{rate1} 
R\simeq\frac{\Gamma^2}{2 \pi c r_e^2}\simeq 2.6\times 10^{8} cm^{-3} s^{-1}.
\end{equation} 

For the inclined field lines applied to jet-like structures, in Fig. 3 we show the dimensionless time scale versus $\theta$. The set of parameters is: $M_8 = 1$, $L_{42}=1$, $r = R_{lc}/\sin\theta$, $\gamma_0 = 1$, $\gamma_{e,p} = 2$. As it is clear from the plot, the instability is still efficient inside the jet structures. But the time-scale when the balance takes place (see Eq. (\ref{bal})) exceeds the kinematic time-scale by several times, indicating the irrelevance of the mentioned process inside the jet-like structures.

For the obtained value of the pair number density (see Eq. (\ref{dens})), the annihilation time-scale becomes of the order of $t_{ann}\simeq1/(\sigma_T n_{\pm}c)\simeq 400$ sec. These pairs, initially mildly relativistic, will be characterised by the synchrotron cooling timescale, $t_{syn}\simeq\gamma mc^2/P_s\simeq 10^7$ sec, exceeding by many orders $t_{ann}$, indicating very low efficiency of the synchrotron process. On the other hand, the particles can centrifugally accelerate, which inevitably reduces the cooling timescale, and the synchrotron process might become important. Therefore, one should consider this scenario as well.

As it has been shown in \cite{review}, the acceleration time-scale of particles is given by
\begin{equation}
\label{tacc} 
t_{acc}\simeq\frac{R_{lc}}{2c}\times\left(1-\frac{r_0^2}{R_{lc}^2}\right)^{1/2},
\end{equation} 
where $r_0$ indicates the initial coordinate of the particle and we have assumed that $\gamma_{pairs}\simeq 1$. By taking into account $\Delta r\simeq\gamma_0R_{lc}/(2\gamma_{e,p})$ one can straightforwardly show that $t_{acc}$ is of the order of $10^4$ sec. This means that the synchrotron mechanism, after the pairs are accelerated still will be inefficient. A similar conclusion comes from the IC scattering, because normally this process is efficient for relativistic electrons, and consequently the particles should be centrifugally energized. But as we have already seen, the acceleration timescale is too large to make the mechanism efficient enough.

\begin{figure}
  \resizebox{\hsize}{!}{\includegraphics[angle=0]{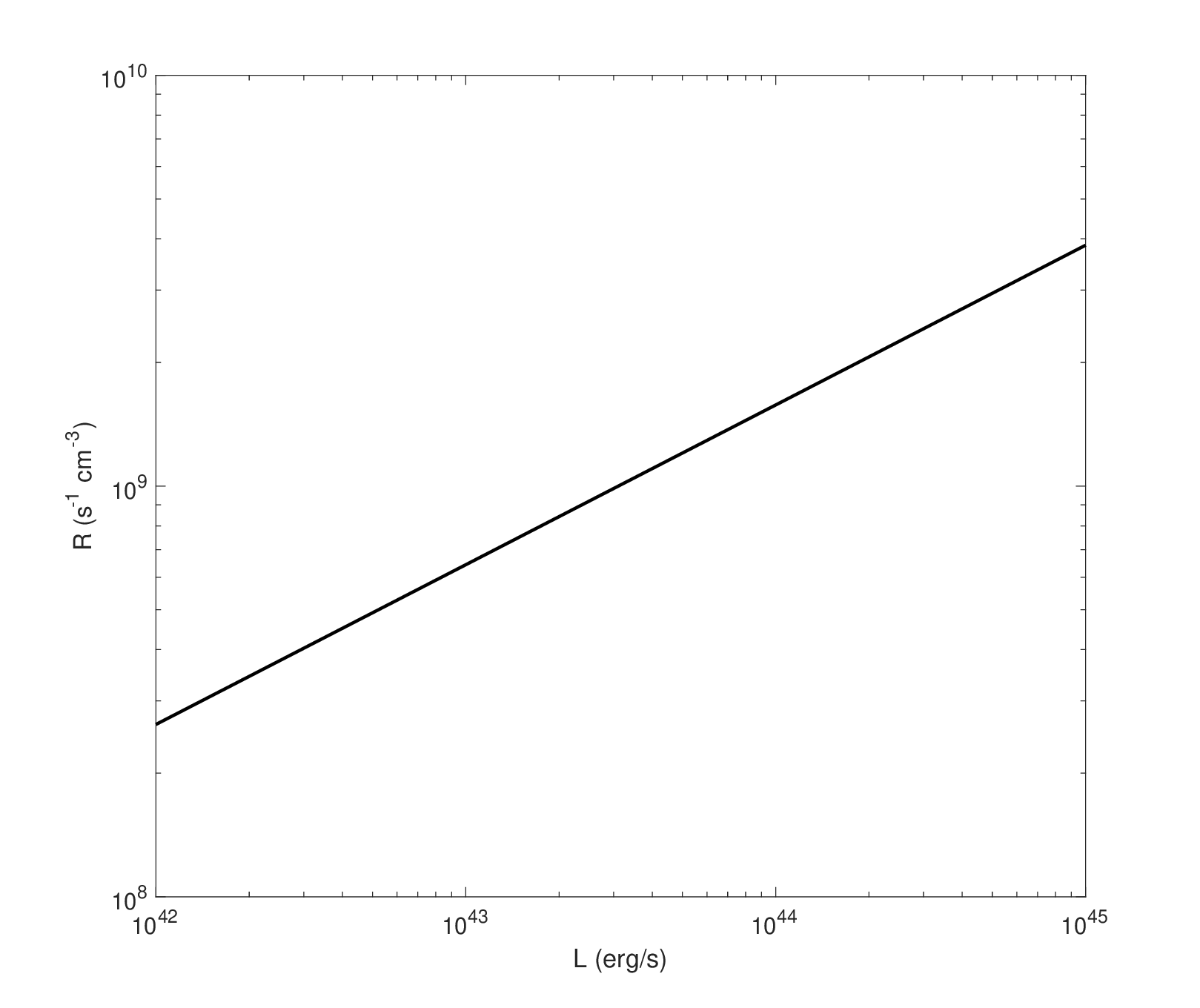}}
  \caption{Pair creation/annihilation rate is shown as a function of $L$. The set of parameters is the same as in Fig. 1 except $\gamma_{e,p} = 10$ and the luminosity.}\label{fig4}
\end{figure}
\begin{figure}
  \resizebox{\hsize}{!}{\includegraphics[angle=0]{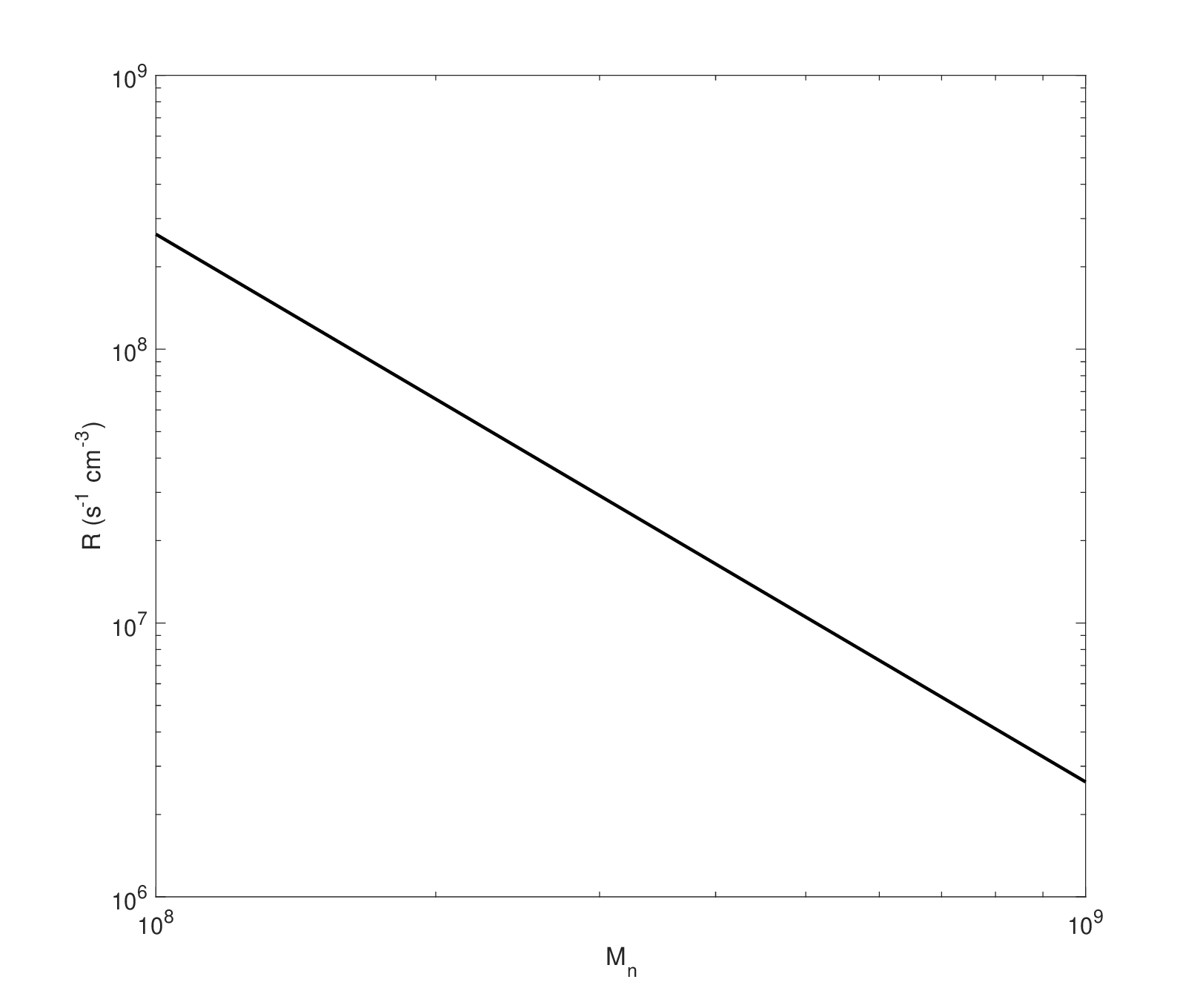}}
  \caption{Pair creation/annihilation rate is shown as a function of $M_n$. The set of parameters is the same as in Fig. 3 except $L = 10^{42}$ erg/s and the range of the black hole mass.}\label{fig5}
\end{figure}

In Fig. 4 we show the pair creation/annihilation rate versus the AGN luminosity. The set of parameters is the same as in Fig. 2 except $\gamma_{e,p} = 10$ and the luminosity range. As it is evident, the higher the luminosity, the higher the pair creation/annihilation rate, which is a natural behaviour.

In Fig. 5, similar behavior has been shown, but versus the normalized black hole mass, $M_n = M/M_{\odot}$. The set of parameters is the same as in Fig. 4, except $L = 10^{42}$ erg/s and the range of the black hole mass. The plot is a continuously decreasing function of the black hole mass, which is a natural result of Eq. (\ref{rotat}): higher the mass, less the angular velocity of rotation, and consequently, the less the centrifugal effects.

This study shows that centrifugally energized AGN magnetospheres should be characterized by the annihilation lines, $2mc^2\simeq 1$ MeV, which might be interesting in the context of multi-wavelength observations of AGN \cite{abdo1,abdo2,aleksic}. Moreover, in AGN astronomy, it is well known that the X-ray and GeV-TeV gamma ray skies have been explored in detail, while the study in the MeV range is not that rich \cite{mev}, therefore, the present study is significant. It is worth noting that in general, this emission will be redshifted because it has an extragalactic origin, and consequently, the observed energy, $\epsilon$, will be reduced by the factor, $1+z$, where $z$ is a redshift of the AGN
\begin{equation}
\label{Eobs} 
\epsilon_{obs} = \frac{1\; MeV}{1+z}.
\end{equation} 
This means that for small redshift AGNs, the annihilation line is of the order of $1$ MeV, whereas for higher redshifts, the energy might be even much less. For example, according to the catalog \cite{highz}, the highest observed redshift is $\sim 6$, implying that the observed annihilation line will be of the order of $140$ keV.

\section{Conclusions}

For the wide range of AGN luminosities and masses, we have studied the efficiency of centrifugally induced pair production.

In particular, by assuming that the AGN magnetosphere is composed of protons and electrons, which are centrifugally energized, we have found that the centrifugal force efficiently induces the exponentially amplifying electrostatic field, which in due course of time reaches the Schwinger threshold, leading to pair production.

It has been shown that the process is balanced by the annihilation mechanism, leading to a saturated value of production/annihilation rate, which has been explored versus the AGN luminosity and the central black hole mass.

We have found that for a wide range of parameters, the mechanism is very efficient, except for the field lines with small inclination angles with respect to the rotation axis, excluding the AGN jets from a class of objects where the studied process is significant.

\section{Acknowledgments}
Z.O. would like to thank  Dr. Fabrizio Tavecchio for interesting comments. The work was supported by the EU fellowships for Georgian researchers, 2023 (57655523). Z.O. also would like to thank Torino Astrophysical Observatory and Universit\'a degli Studi di Torino for hospitality during working on this project.

%%%%%%%%%%

\end{document}